\documentclass [11pt,twoside]{article}

\usepackage{shrthnds}
\usepackage{cite}
\usepackage{graphicx}
\usepackage{color}
\usepackage{subfigure}
\usepackage{setspace}
\usepackage{yfonts}[1988/10/03]
\usepackage{multirow}
\def\be{\begin{equation}}
\def\ee{\end{equation}}
\def\ba{\begin{eqnarray}}
\def\ea{\end{eqnarray}}

\usepackage[hang,small]{caption}

\usepackage{geometry}
\usepackage{fancyhdr}
\usepackage{titlesec,titletoc}
  \titleformat{\section}{\Large\sf\bfseries}{\thesection}{1em}{}
  \titleformat{\subsection}{\large\sf\bfseries}{\thesubsection}{1em}{}

\title{\sf\bfseries Reissner Nordstrom Metric in Unimodular Theory of Gravity}
\author{\normalsize  Pankaj Chaturvedi$^1$\footnote{email: cpankaj@iitk.ac.in}~, 
 Naveen K. Singh$^2$\footnote{email: naveensinghctp@gmail.com}~ 
 and Dharm Veer Singh$^2$\footnote{email: veerdsingh@gmail.com}}
\date{}

\begin{document}
\maketitle
\vspace{-0.6cm}
\bc
{\small  $^1$Department
 of Physics, Indian Institute of Technology Kanpur, \\
 Kanpur 208 016, India
 \\}
{\small  $^2$Centre for Theoretical Physics, Jamia Millia Islamia, \\
New Delhi-110025, India\\
}
\ec\texttt{}

\centerline{\small\date{\today}}
\vspace{0.5cm}

\bc
\begin{minipage}{0.9\textwidth}\begin{spacing}{1}{\small {\bf Abstract:}
We study the modified  Reissner Nordstrom metric in the unimodular gravity. So far the spherical symmetric
Einstein field equation in unimodular gravity has been studied in the
absence of any source. We consider static electric and magnetic charge as source. We solve for Maxwell equations in unimodular gravitational
 background. We show that in unimodular gravity the electromagnetic field strength tensor is modified. We also show that the solution in unimodular 
gravity differs from the usual R-N metric in Einstein gravity with some corrections. We further study the thermodynamical
properties of the R-N black-hole solution in this theory.

}\end{spacing}\end{minipage}\ec

\vspace{0.5cm}\begin{spacing}{1.1}


\section{Introduction}
The unimodular gravity is an alternative theory for the cosmological constant.  Initially it was introduced by 
Einstein \cite{Einstein}. It was further  developed in  the Ref. \cite{Anderson}. The concept of cosmological constant was 
introduced by Einstein to explain the static universe. However, the idea of static universe was discarded by Einstein
after Hubble's experiment in year 1929. The cosmological constant was not paid attention till the year 1967
\cite{PSS,Shk,R_R} when the loitering in the universe's expansion was observed and to support this  
a positive cosmological constant was considered. The great discovery in year 1998 \cite{Perlmutter,Riess}, which
confirmed that the universe is accelerating, gave the birth of many theories such as quintessence, K-essence,
phantom field based model, Chaplygin gas model, cosmological constant based theories etc. to describe it. The unimodular gravity was rethought, since the cosmological constant is its inherent property which provides
an explanation  for the accelerating universe.
In this theory, the determinant of metric is not dynamical and hence in Einstein-{\it unimodular} equation the cosmological
constant is not present, however it appears naturally as an integration constant \cite{Weinberg}. Since, the cosmological
constant is not present in Einstein-{\it unimodular} equation, unimodular gravity may solve the fine tuning problem
present in the cosmological constant.

One another beautiful feature of unimodular gravity is that it can explain
the current acceleration of the universe by assuming single component \cite{Jain:2011,Jain:2012gc} such as either only cosmological constant, or only non-relativistic
matter. In that proposal of  Refs \cite{Jain:2011,Jain:2012gc}, the metric is first decomposed into a scalar field and unimodular metric, where
the scalar field explain the universe dynamics and unimodular metric satisfies the unimodular constraint. The gravitational
action is written in terms of unimodular metric and the scalar field. The full general covariance is then broken
by introducing a parameter. The action is now invariant under only {\it unimodular}-general coordinate transformations.
Considering this theory, with the new parameter, one may show that either only cosmological constant or only
non-relativistic matter can give rise the acceleration in the universe. Considering this proposal, the inflation is
also studied \cite{Cho:2014taa}. We point out that this proposal slightly differs from the standard unimodular theory of
gravity where no such parameter is introduced. The perturbation analysis of standard unimodular gravity
is studied in Refs. \cite{Bradenberger,Basak}.  The unimodular gravity is further generalized recently in Refs. 
\cite{Bamba:2016wjm,Nojiri:2016plt,Odintsov:2016imq,Nassur:2016yhc,Nojiri:2016ppu,Nojiri:2016ygo,Nojiri:2015sfd }. Some 
progress and applications of unimodular theory have also been addressed in the Refs. 
\cite{abbassi1,Weinberg,Fiol_Garriga,abbassi2,Bock,Alvarez,Alvarez_Faedo,Farajollahi,Ng_Dam,FGB,Earman}. In this paper, we consider
the standard unimodular gravity. 

On the other hand the R-N metric gives the geometry of  space time around the non rotating charged black hole.  R-N metric's solution 
is theoretically interesting and applicable for short time within which the black hole remains charged by some
perturbations like gravitational collapse. In present work we give the details of derivation for modified R-N metric 
in the presence of unimodular gravity. The unimodular constraint is given by,
 \ba
 \sqrt{-g'} =1 . \label{uniconst}
 \ea
 Here, in the metric $g'_{\mu\nu}$, the components of metric are dynamical. However, these components satisfy
 the unimodular constraint (\ref{uniconst}). The unimodular gravity is therefore falls in subclass of general 
 relativity with reduced  degree of freedom \cite{Weinberg}. The standard form of Einstein's equation with cosmological constant is given by,
 \ba
 R'_{\mu\nu} - \frac{1}{2} g'_{\mu\nu} R' + \Lambda g'_{\mu\nu} = 8 \pi G T_{\mu\nu},
 \ea
 where $R'$ and $R'_{\mu\nu}$ are Ricci scalar and Ricci tensor respectively. Now we apply the constraint $\sqrt{-g'}=1$, i.e., $g'_{\mu\nu}\delta g'^{\mu\nu}=0$ which eliminates all the terms
 those are proportional to $g'_{\mu\nu}$. Thus, by taking the trace and eliminating those terms, we have,
\ba
\left(R'_{\mu\nu}-\frac{1}{4} g'_{\mu\nu} R'\right)=8 \pi G  \left( T_{\mu\nu} -\frac{1}{4} g'_{\mu\nu}T\right). \label{unimodequation}
\ea
This equation contains the information $\sqrt{-g'}=1$ and so called; unimodular gravity equation. 

The paper is organized as follows. In Sec. (\ref{sec_uni_rad}), we derive the Einstein-unimodular equation considering the charge 
as a source. In Sec. (\ref{Maxwell}), we solve for the Maxwell equation and in Sec. (\ref{E_Uni}), Ricci tensor, Ricci scalar and energy-
momentum-tensor are written in unimodular gravity and we further solve for the metric by writing all the components
of Einstein's unimodular equations. Thermodynamics is studied in Sec. (\ref{THERMO}). Finally we conclude in the last section.
\section{ Einstein's unimodular equation in the presence of radiation} \label{sec_uni_rad}
We scale the metric such that the determinant of metric remains same as in the case of Minkowski space, i.e., after transformation in the cartesian coordinates
the determinant  is unity and it becomes $r^2 \sin(\theta)$ in the spherical coordinates. Let us consider a metric $g_{\mu\nu}$ given as,

\ba
g_{\mu\nu}= [B(r), -A(r),- r^2, -r^2 \sin^2{(\theta)}].
\ea
Therefore, the required scaling 
parameter is $\left(A(r)B(r)\right)^{1/4}$ and the scaled metric $g'_{\mu\nu}$ may written as
\ba
g'_{\mu\nu} = \frac{g_{\mu\nu}}{\left(A(r)B(r)\right)^{1/4}} \label{scaled_metric}
\ea

 The condition (\ref{uniconst}) changes the definition of energy momentum tensor, since $\delta(- g')^{1/2}=0$. For the standard Lagrangian for electromagnetic field,
\ba
L_{rad}=\frac{1}{4}F_{\mu\nu}F^{\mu\nu},
\ea 
the energy momentum tensor in unimodular gravity turns out to be
\ba
T_{\mu\nu}=g'^{\rho\sigma} F_{\rho\mu}F_{\sigma\nu}.
\ea
Including the action corresponding the electromagnetic field, the modified Einstein's unimodular equation can be written as
\ba
\left( R'_{\mu\nu}-\frac{1}{4} g'_{\mu\nu} R'\right)=8 \pi G \left(g'^{\rho\sigma} F_{\rho\mu}F_{\sigma\nu} -\frac{1}{4} g'_{\mu\nu}T\right),
\ea
where, T is  trace of the energy momentum tensor of radiation and $R'_{\mu\nu}$ is scaled Ricci tensor.

\section{Maxwell Equations:} \label{Maxwell}
\vspace{-.2cm}
In this section, we solve for all the components of field strength tensor by using Maxwell's equations.
The electric field strength tensor may be written as \cite{Mammadov}

\[ F_{\mu\nu} = \left( \begin{array}{cccc}
0& f(r,t)&0 &0 \\
-f(r,t)& 0 &0 &0 \\
0& 0&0 &r^2 g(r,t)Sin\theta\\
0& 0& -r^2 g(r,t)Sin\theta& 0 \end{array} \right)\] 
and corresponding Maxwell's equation are given by,
\ba
g'^{\mu\nu}\nabla_\mu F_{\nu\rho} =J_\nu,\label{max_1}
\ea
and
\ba
\nabla_{[\mu} F_{\nu\rho]} =0,\ \ or \ \ \partial_{ [\mu} F_{\nu\rho]} =0 \label{max_2} ,
\ea
where all covariant derivatives are denoted by $\nabla$. Since we have not considered the current, the spatial part of
equation (\ref{max_1}) gives us

\ba
\frac{1}{B(r)}\frac{\partial f(r,t)}{\partial t} = 0 .\label{evol_eq2}
\ea
The evolution equation (\ref{evol_eq2}) implies that the function $f(r,t)$ is time independent.
\ba
f(r,t) = f(r)\label{time_ind}
\ea 
Using the equation (\ref{time_ind}) and applying Gauss's law from the time component of (\ref{max_1})  we get
\ba
f(r,t) = f(r) = \frac{Q \sqrt{A(r) B(r)}}{4 \pi r^2} \label{electric}.
\ea
 On the other hand the
equation (\ref{max_2}) gives the relation for  the magnetic field of theoretical magnetic charge P without any modification.
\ba
g(r,t)=g(r)= \frac{P}{4 \pi r^2} \label{magnetic}.
\ea
\section{Einstein-Unimodular Equation} \label{E_Uni}
In this section, we write Ricci scalar $R'$ and Ricci tensor $R'_{\mu\nu}$ in terms of new unimodular metric $g'_{\mu\nu}$.
The new Christoffel symbols $\Gamma'^{\alpha}_{\mu\nu}$ are associated to the unimodular metric $g'_{\mu\nu}$.
\subsection{Curvature Scalar:}
In the calculation of $R'$ we use the new metric $g'_{\mu\nu}$ given by the Eq (\ref{scaled_metric}) .
We may evaluate the Riemann curvature and Ricci tensors respectively by following relations:
\ba 
& R'^\rho_{\sigma\mu\nu}= &\partial_\nu\Gamma'^\rho_{\mu\sigma}-\partial_\mu\Gamma'^\rho_{\nu\sigma}
+\Gamma'^\rho_{\nu\lambda}\Gamma'^\lambda_{\mu\sigma}-
\Gamma'^\rho_{\mu\lambda}\Gamma'^\lambda_{\nu\sigma} \label{curv_tensor}\\
& R'_{\sigma\nu}= &\partial_\nu\Gamma'^\rho_{\rho\sigma}-\partial_\rho\Gamma'^
\rho_{\sigma\nu}+\Gamma'^\rho_{\nu\lambda}\Gamma'^\lambda_{\rho\sigma}
-\Gamma'^\rho_{\rho\lambda}\Gamma'^\lambda_{\nu\sigma}, \label{ricci_tensor} 
\ea
 where, $\Gamma'^{\mu}_{\alpha \beta}$ is described in the Refs. \cite{abbassi1,abbassi2} in details. The components of Eq. (\ref{ricci_tensor}) results
\ba
R'_{tt} &=&-\frac{B(r) A(r)'}{4 r A(r)^2} +\frac{7 B(r) (A(r)')^2}{32 A(r)^3} + \frac{3 B(r)'}{4 r A(r)} \nonumber \\
&-&\frac{5 B(r)'A(r)'}{16 A(r)^2}-\frac{9 (B(r)')^2}{32 A(r) B(r)} - \frac{ B(r) A(r)''}{8 A(r)^2}
+\frac{3 B(r)''}{8 A(r)} ,\label{ricci_1} \\
R'_{rr}&=&\frac{5 A(r)'}{4 r A(r)} - \frac{9  (A(r)')^2}{16 A(r)^2} + \frac{ B(r)'}{4 r B(r)} + \frac{ B(r)'A(r)'}{8 A(r) B(r)}-\frac{ (B(r)')^2}{16 B(r)^2}
+\frac{3 A(r)''}{8 A(r)} - \frac{ B(r)''}{8 B(r)}, \label{ricci_2} \\
R'_{\theta \theta}&=&1- \frac{1}{ A(r)} + \frac{r  A(r)'}{ A(r)^2} - \frac{ 7 r^2 (A(r)')^2}{32 A(r)^3} - \frac{ r^2 B(r)'A(r)'}{16 A(r)^2 B(r)}\nonumber \\
&-&\frac{3 r^2 (B(r)')^2}{32 A(r) B(r)^2} +\frac{r^2 A(r)''}{8 A(r)^2} + \frac{r^2 B(r)''}{8 A(r) B(r)},\label{ricci_3} \\
 R'_{\phi\phi}&=& \sin^2{\theta}R'_{\theta\theta},\label{ricci_4}
\ea
where $A(r)$ and $B(r)$ are defined by the line element in usual gravity:
\ba
ds^2=g_{\mu\nu}dx^{\mu}dx^{\nu}=B(r)dt^2-A(r)dr^2-r^2(d\theta^2+sin^2\theta d\phi^2). \label{METRIC}
\ea
From equation (\ref{ricci_1}) to (\ref{ricci_4}) we can write Ricci scalar  $R'$ as 
\ba
R' = g'^{\mu\nu}R'_{\mu\nu},
\ea
which turns out to be
\ba
R' &=&\left( A(r) B(r)\right)^{1/4} \Big[-\frac{2}{r^2} +\frac{2}{r^2 A(r)} -\frac{7 A(r)'}{2 r A(r)^2} +\frac{39 (A(r)')^2}{32 A(r)^3} \nonumber\\
&+&\frac{B(r)'}{2 r A(r) B(r)} -\frac{5 A(r)' B(r)'}{16 A(r)^2 B(r)} -\frac{(B(r)')^2}{32 A(r) B(r)^2} -\frac{3 A(r)''}{4 A(r)^2} + \frac{B(r)''}{4 A(r) B(r)}\Big] \\
&=& \left(A(r) B(r)\right)^{1/4} R \label{eqn_R},
\ea
where R is defined as
\ba
R&=&\Big[-\frac{2}{r^2} +\frac{2}{r^2 A(r)} -\frac{7 A(r)'}{2 r A(r)^2} +\frac{39 (A(r)')^2}{32 A(r)^3} \nonumber\\
&+&\frac{B(r)'}{2 r A(r) B(r)} -\frac{5 A(r)' B(r)'}{16 A(r)^2 B(r)} -\frac{(B(r)')^2}{32 A(r) B(r)^2} -\frac{3 A(r)''}{4 A(r)^2} + \frac{B(r)''}{4 A(r) B(r)}\Big],
\ea
which is computed with respect to the original metric $g_{\mu\nu}$.
%
The components of energy momentum tensor for radiation may be written as,
\ba
T_{tt}&=&-\left(A(r)B(r) \right)^{1/4}\frac{f(r)^2}{A(r)}, \nonumber \\
T_{rr}&=&\left( A(r)B(r) \right)^{1/4}\frac{f(r)^2}{B(r)}, \nonumber \\
T_{\theta\theta}&=& -r^2 g(r)^2 \left( A(r)B(r) )\right)^{1/4},  \nonumber \\
T_{\phi\phi} &=& -r^2 g(r)^2 \sin^2(\theta)\left( A(r)B(r) \right)^{1/4}.
\ea
Substituting all the values of $R'_{\mu\nu}$, $R'$ and energy momentum tensor in Eq. (\ref{unimodequation}), we obtain all the components
(tt, rr, $\theta \theta$) of 
equations as follows,
\ba
&&\frac{B(r)}{2r^2}-\frac{B(r)}{2r^2 A(r)} - 4 \pi G \left(A(r)B(r) \right)^{1/4}\Bigg[\frac{f(r)^2 }{A(r)}+ B(r) g(r)^2\Bigg] +\frac{5 B(r) A(r)'}{8 r A(r)^2} -\frac{11 B(r) (A(r)')^2}{128 A(r)^3}
\nonumber\\
&&+\frac{5 B(r)'}{8 r A(r)} -\frac{15 A(r)' B(r)'}{64 A(r)^2}
-\frac{35 (B(r)')^2}{128 A(r) B(r)} +\frac{B(r) A(r)''}{16 A(r)^2} + \frac{5 B(r)''}{16 A(r)} =0 , \label{Einstein1} \\ 
&&\frac{1}{2r^2}-\frac{A(r)}{2r^2} + 4 \pi G \left(A(r)B(r) \right)^{1/4}\Bigg[\frac{f(r)^2 }{B(r)} + A(r) g(r)^2 \Bigg]+\frac{3  A(r)'}{8 r A(r)} -\frac{33  (A(r)')^2}{128 A(r)^2} \nonumber \\
&&+\frac{3 B(r)'}{8 r B(r)} +\frac{3 A(r)' B(r)'}{64 A(r) B(r)}
-\frac{9 (B(r)')^2}{128  B(r)^2} +\frac{3 A(r)''}{16 A(r)} - \frac{ B(r)''}{16 B(r)} =0 ,\label{Einstein2} \\
&&\frac{1}{2}-\frac{1}{2 A(r)} - 4 \pi G r^2 \left(A(r)B(r) \right)^{1/4} \Bigg[\frac{  f(r)^2}{A(r) B(r)} + g(r)^2 \Bigg]  + \frac{r  A(r)'}{8 A(r)^2} +\frac{11 r^2 (A(r)')^2}{128 A(r)^3}\nonumber \\
&&+\frac{r B(r)'}{8 A(r) B(r)} -\frac{9 r^2 A(r)' B(r)'}{64 A(r)^2 B(r)}
-\frac{13 r^2 (B(r)')^2}{128 A(r) B(r)^2} -\frac{r^2 A(r)''}{16 A(r)^2} + \frac{3 r^2 B(r)''}{16 A(r) B(r)} =0, \label{Einstein3} 
\ea
and $\phi \phi$ component gives same equations as  $\theta \theta$. Now multiplying $\frac{A(r)}{B(r)}$ in equation (\ref{Einstein1}) and adding with (\ref{Einstein2}) or multiplying $\frac{A(r)}{r^2}$ in equation (\ref{Einstein3}) and adding with
 (\ref{Einstein2}) results,
\ba
\frac{A(r)'}{ r A(r)} -\frac{11 (A(r)')^2 }{32 A(r)^2} +\frac{B(r)'}{r B(r)} -\frac{3 B(r)' A(r)'}{16 A(r) B(r)} - \frac{11 (B(r)')^2}{32 B(r)^2} +
\frac{A(r)''}{4 A(r)} + \frac{B(r)''}{4 B(r)} =0 . \label{eq:35} 
\ea
Adding the equation (\ref{Einstein2}) and multiplication of $\frac{3 A(r)}{r^2}$ with equation (\ref{Einstein3}) implies,
\ba
\frac{B(r)''}{B(r)} &=& \frac{2}{r^2} -\frac{2 A(r)}{r^2} +16 \pi G \left(A(r)B(r) \right)^{1/4} \Bigg[\frac{ f(r)^2}{ B(r)} + A(r) g(r)^2\Bigg]-\frac{3 A(r)'}{2 r A(r)} \nonumber \\
&-&\frac{3 B(r)'}{2 r B(r)} + \frac{3 A(r)' B(r)'}{4 A(r) B(r)}
+\frac{3 (B(r)')^2}{4 B(r)^2} . \label{eq:B2}
\ea
We get a relation between $A(r)$ and $B(r)$ from equation (\ref{eq:35});
\ba
A(r) = \frac{1}{B(r)}\left[1+\frac{C_2}{8 r^3}\right]^{-8/3}\label{eq:A_1}.
\ea 
From equation (\ref{eq:B2}) and (\ref{eq:A_1}) we have,
\ba
B(r)''&=& -\frac{2}{r^2}\left(1+\frac{C_2}{8 r^3}\right)^{-8/3} -\frac{3 C_2 B(r)}{2 r^5}\left(1+\frac{C_2}{8 r^3}\right)^{-1} + \frac{2 B(r)}{r^2} \nonumber \\
      &+& \frac{G \left(A(r)B(r) \right)^{5/4}}{\pi r^4}\Bigg[Q^2+ P^2\Bigg]  +\frac{3 C_2 B(r)'}{4 r^4}\left(1+\frac{C_2}{8 r^3}\right)^{-1}.
\ea
Here we have used the equations (\ref{electric}) and (\ref{magnetic})  for $f(r)$ and $g(r)$ respectively. The solution of differential equation of $B(r)$ 
may be written  as
\ba
B(r) &=& \frac{4\left(8 + C_2/r^3\right)^{1/3}r^3}{C_2 + 8r^3} +\frac{4G \left(8 + C_2/r^3\right)^{1/3} r }{\pi \left(C_2 + 8r^3\right)}\Bigg[ 
\left(\frac{\left(8+ C_2/r^3\right)^{1/3} r^9}{\left(C_2+ 8r^3\right)^{3}}\right)^{1/4} \left(P^2 + Q^2\right) \Bigg] \nonumber \\
&+& \frac{r^2 C_3}{C_2 + 8 r^3} +\frac{r^5 C_4}{3\left(C_2 + 8 r^3\right)},\label{eq_B_r}
\ea
 where $C_1$, $C_2$ and $C_3$ are constants. The expression for $A(r)$ can be obtained from the equation (\ref{eq:A_1}). Considering the constant $C_2$ very small 
from the Eq. (\ref{eq_B_r}) the value of $B(r)$ can be expanded as
\ba
B(r)&=& 1+ \left(\frac{C_3}{8}-\frac{C_4 C_2}{192}\right)\frac{1}{r} + \frac{G}{4\pi}\frac{P^2+Q^2}{r^2} -\frac{C_2}{12 r^3} -\left( \frac{C_3}{64} -
\frac{C_4 C_2}{1536}\right)\frac{C_2}{r^4} \nonumber \\ 
&-& \frac{G C_2}{24\pi}\left(\frac{P^2+Q^2}{r^5}\right) + \frac{C_4}{24} r^2, \nonumber \\
&=& 1+\frac{G(P^2+Q^2)}{4\pi r^2}+\frac{C_3}{8r}+\frac{C_4 r^2}{24}-C_2\Big(\frac{G(P^2+Q^2)}{24\p r^5}
+\frac{1}{12r^3}+\frac{C_3}{64r^4}+\frac{C_4}{192 r}\Big)+\frac{C_2^2C_4}{1536r^4} , \label{metric_func} \nonumber 
\\
\ea
where we have expanded the term  up to order of $r^{-5}$.
The numerical constants $C_3,\,C_4$ are  related to the  mass $M$ and cosmological constant respectively.
The cosmological constant term $C_4$ appears due to   one of the feature of unimodular theory. However, the constant $C_2$ gives the additional corrections.  The metric defined in Eq. (\ref{METRIC}) describes
the RN black hole solution as in the special case; where $C_2=0$, $C_3=-16 M$ and $C_4=0$  with $(P^2+Q^2) \neq 0$. The
constants $C_2=0$, $C_3=-16M$,  and $C_4=-24 C$ with $(P^2+Q^2)\neq 0$ deals   RN black hole in the presence of cosmological constant.
The constants $C_2 \neq 0$, $C_3=-16M$,  and $C_4=-24 C$ with $(P^2+Q^2)\neq 0$ gives the extra corrections due the non zero value of $C_2$.
For  $C_2=0, \ C_3=-16M,\ (P^2+Q^2)=0$  and $C_4=0$ , the metric reduces to Schawarschild black hole 
solution. In brief, we can describe  three cases as follows; \\
(1.) $C_2=0$, $C_3=-16 M$ and $C_4=0$  with $(P^2+Q^2) \neq 0$     (Standard RN black hole) \\
(2.)  $C_2=0$, $C_3=-16M$,  and $C_4=-24 C$ with $(P^2+Q^2) \neq 0$ (RN black hole in the presence of cosmological constant) \\
(3.)   $C_2 \neq 0$, $C_3=-16M$,  and $C_4=-24 C$ with $(P^2+Q^2)\neq 0$ (RN black hole in the presence of cosmological constant with
the additional corrections).\\

\begin{figure}[h]
\begin{tabular}{c c c c}
\includegraphics[width=0.5\linewidth]{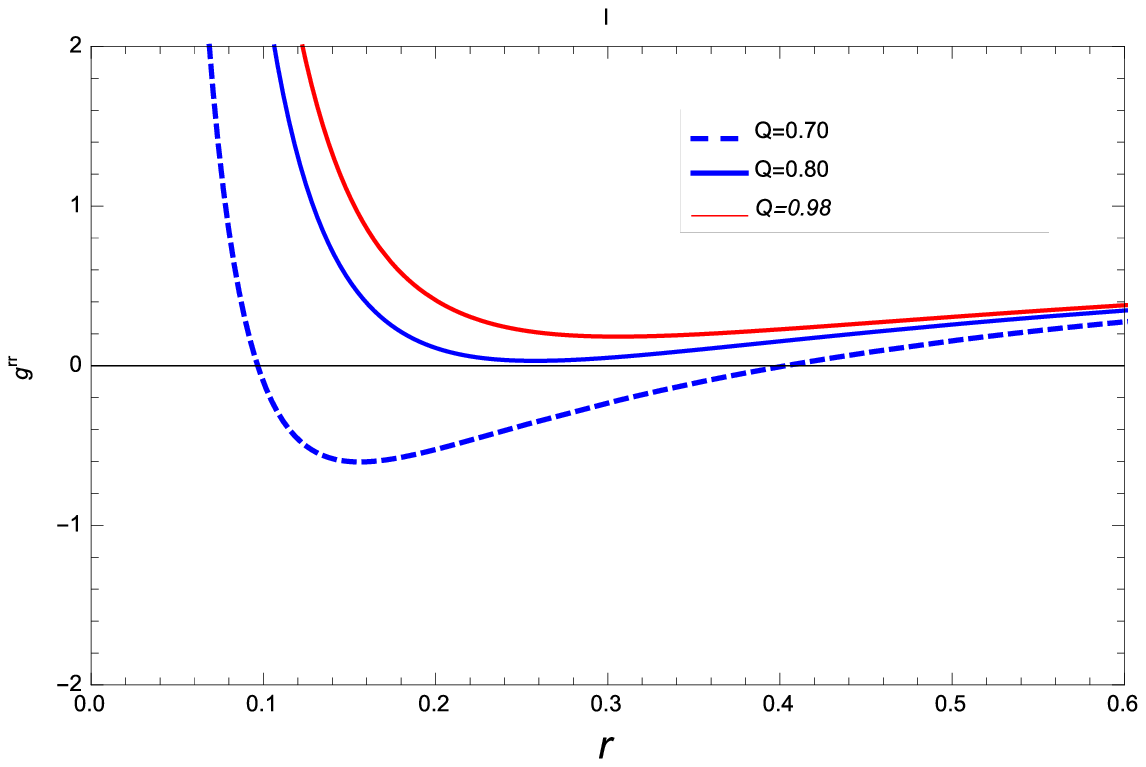}
\includegraphics[width=0.5\linewidth]{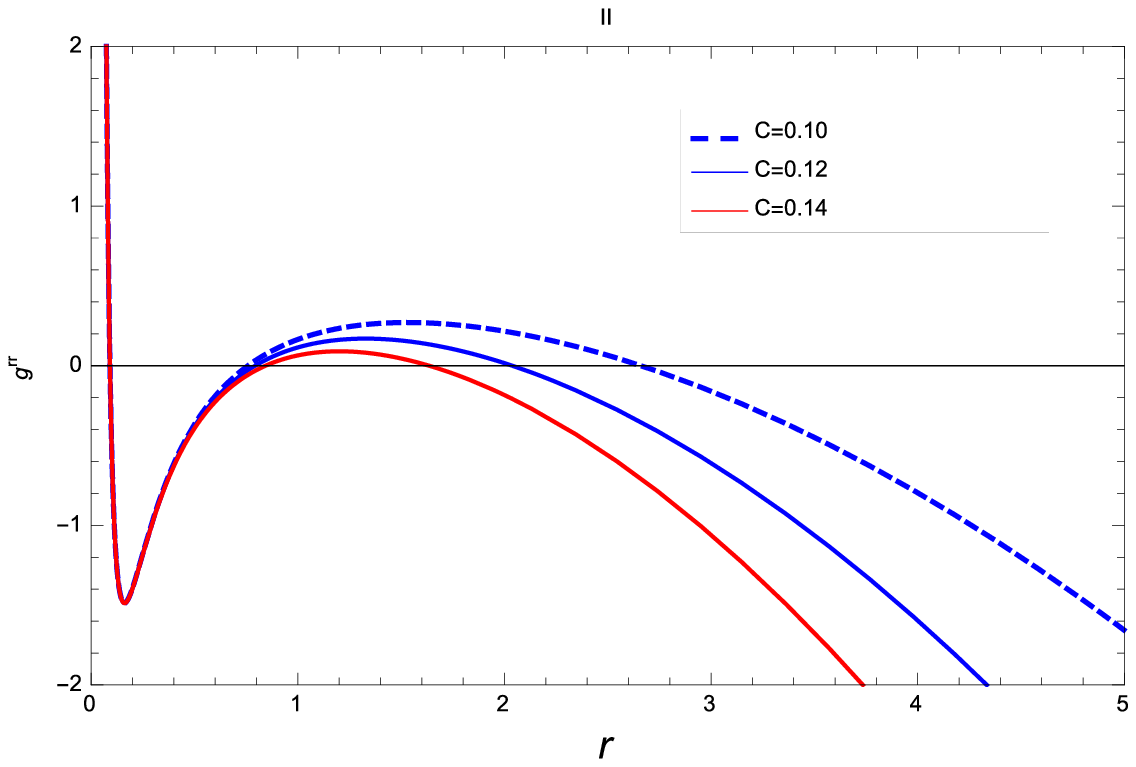}\\
\includegraphics[width=0.5\linewidth]{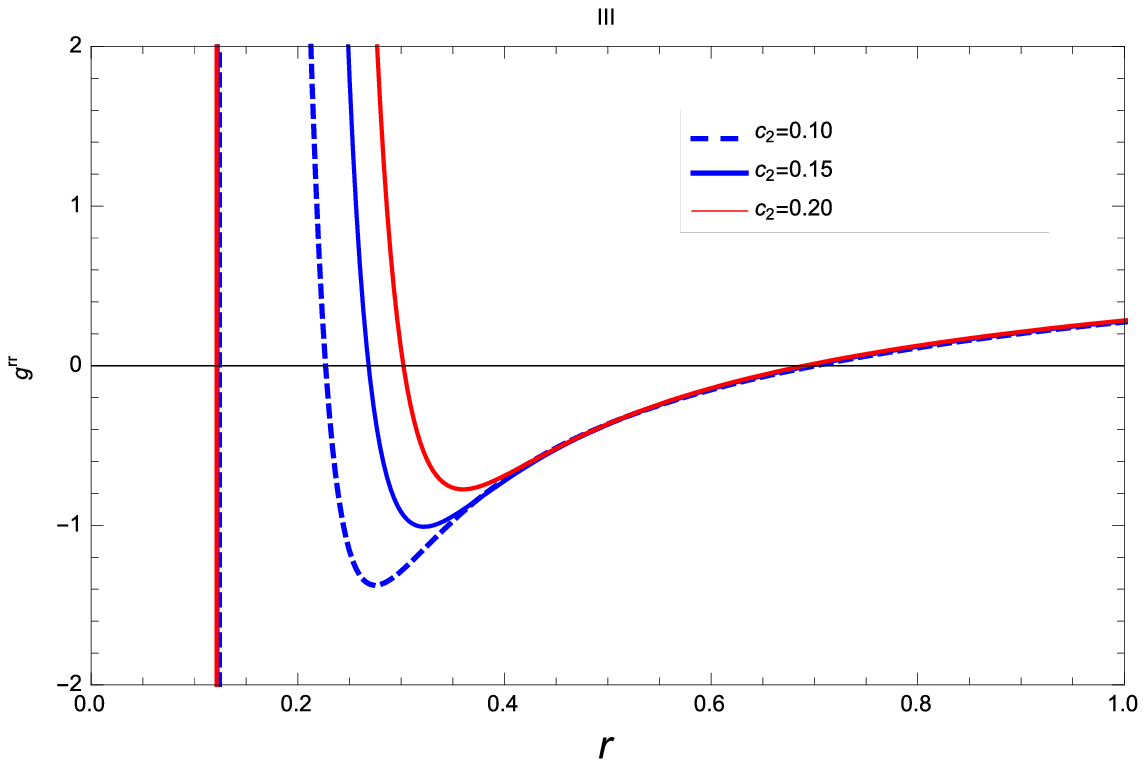}
\includegraphics[width=0.5\linewidth]{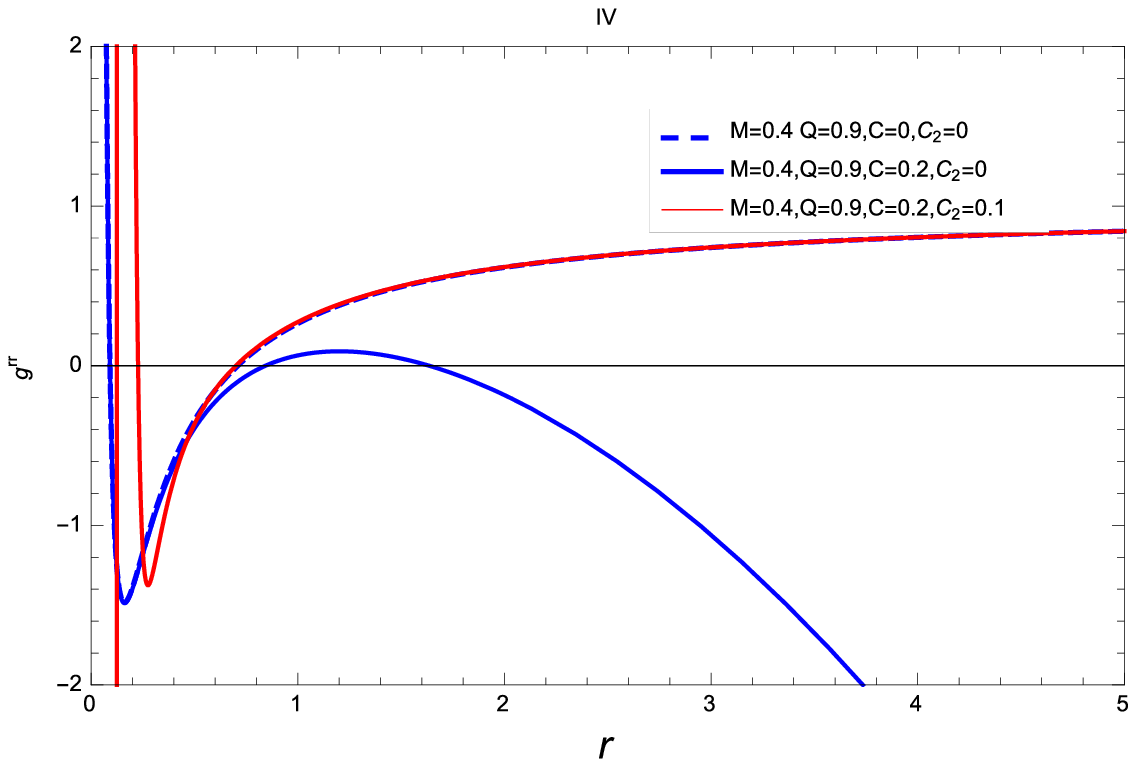}
\end{tabular}
\caption{\label{fig:RK0} The variation  of $g^{rr}$ vs. $r$ is plotted.  The subfigure (I) corresponds the case (1) with $P=0$ as mentioned above. 
The subfigure (II) demonstrate the case (2) with $Q=0.9$; $P=0$ and subfigure (III) is for the case (3) with $Q=0.9$; $P=0$.
The last subfigure (IV) gives a comparative plot for all three cases.}
\end{figure}

The event horizon   can be found by solving the equation, $g^{rr}=0$, i.e.,
\be
\Big(8+\frac{C_2}{r^3}\Big)^{8/3}\Big(1+\frac{P^2+Q^2}{4\pi r^2}+\frac{C_3}{8r}+\frac{C_4 r^2}{24}
-C_2\Big(\frac{P^2+Q^2}{24\p r^5}+\frac{1}{12r^3}+\frac{C_3}{64r^4}+\frac{C_4}{192 r}\Big)+\frac{C_2^2C_4}{1536r^4}\Big)=0 . \nn 
\label{grr}
\ee
The boundaries of the horizon for the cases (1), (2) and (3) are depicted in Fig. (\ref{fig:RK0}).

\section {Thermodynamics } \label{THERMO}
\begin{figure}[h]
\begin{tabular}{c c c c}
\includegraphics[width=0.5\linewidth]{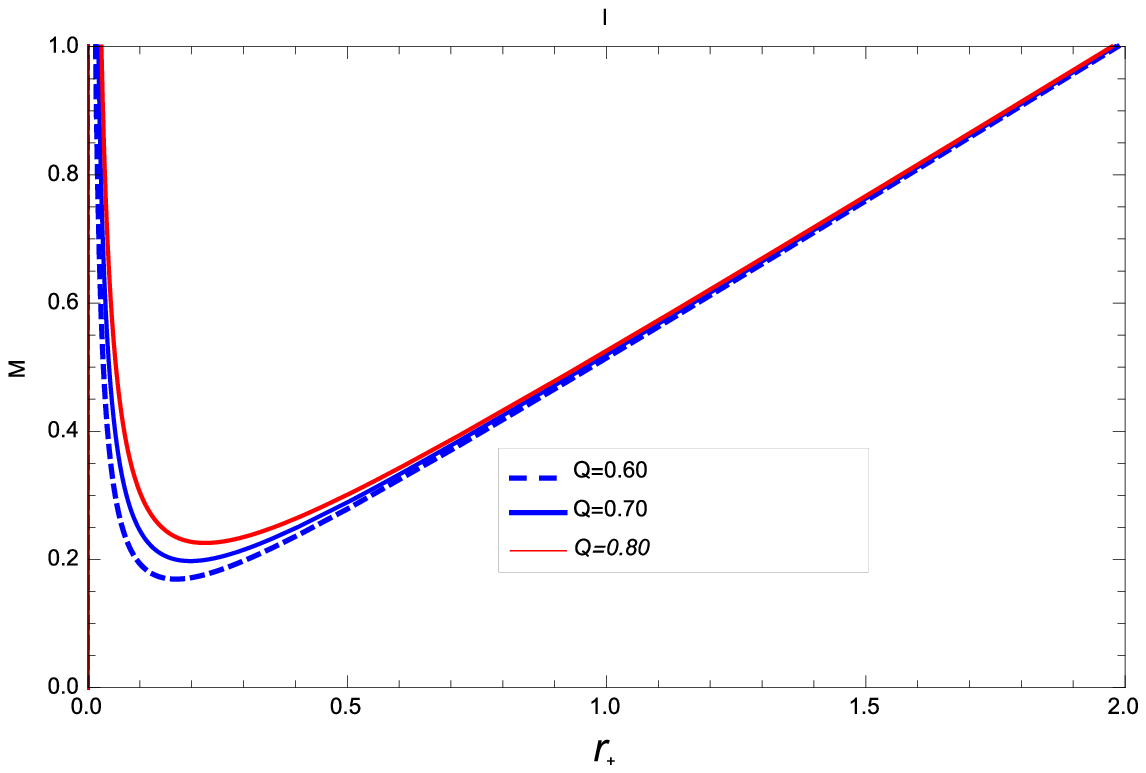}
\includegraphics[width=0.5\linewidth]{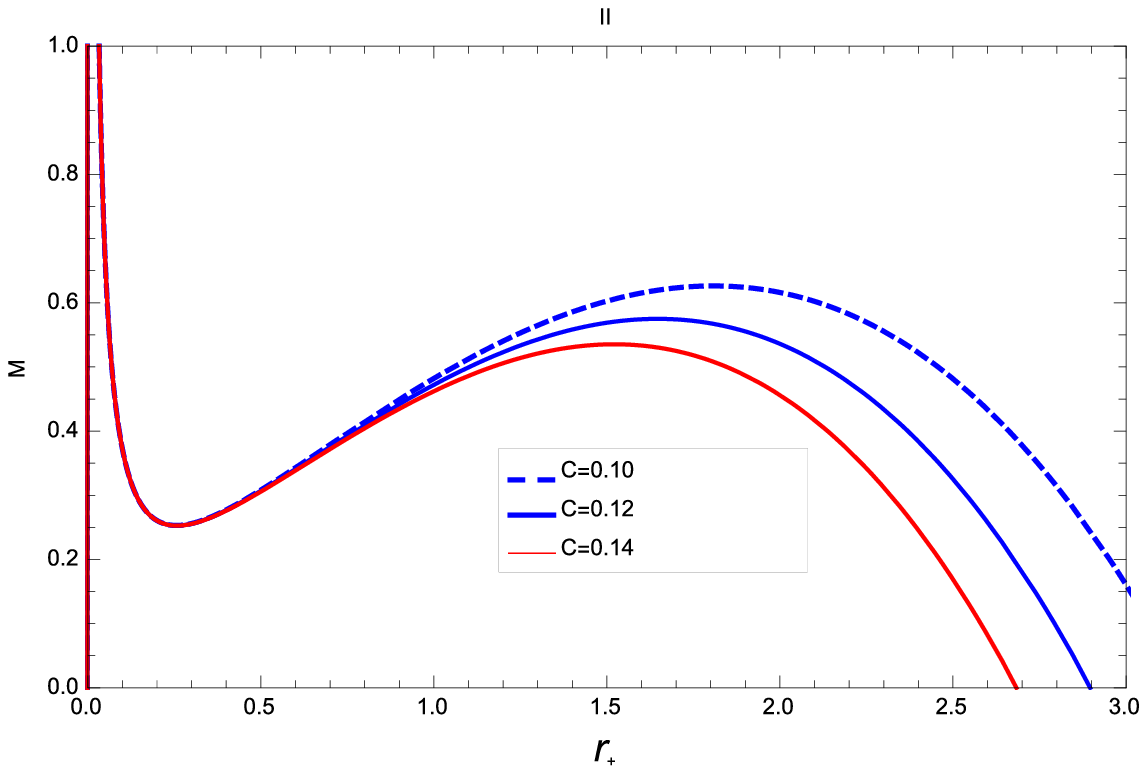}\\
\includegraphics[width=0.5\linewidth]{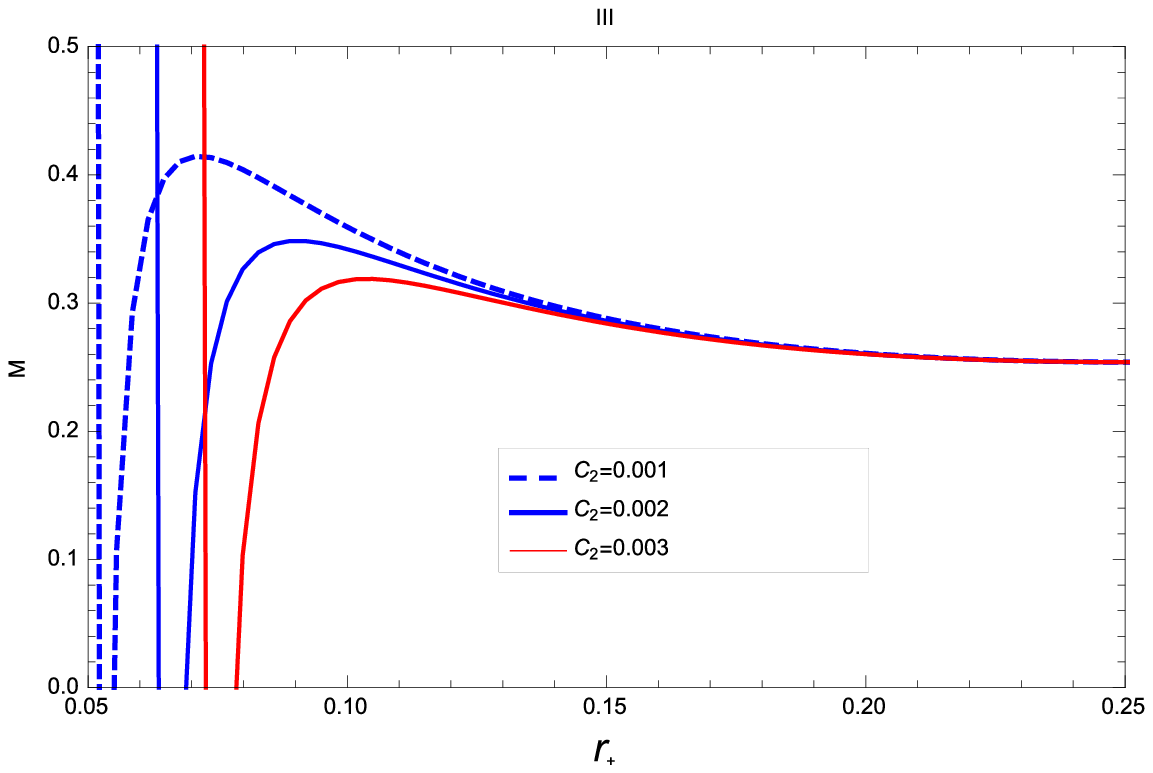}
\includegraphics[width=0.5\linewidth]{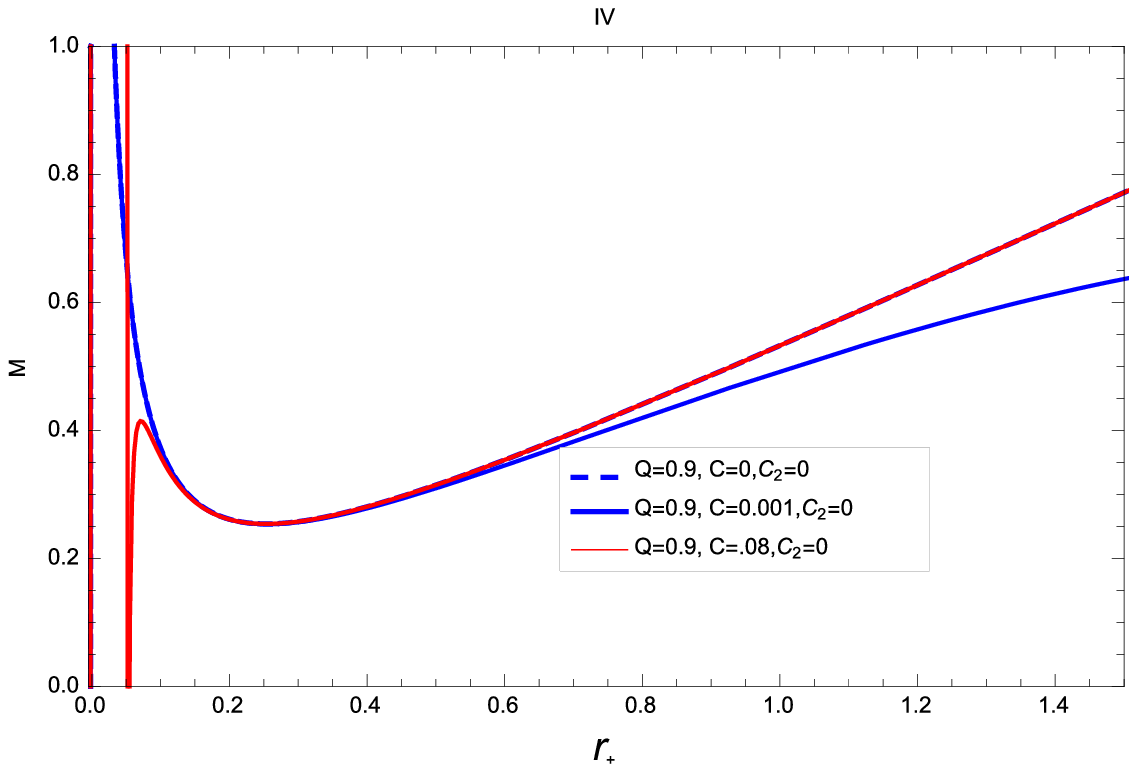}
\end{tabular}
\caption{\label{fig:RK} The plot  of mass with horizon radius $r_+$ is shown in this figure. The subfigure (I) corresponds the case (1)
as mentioned above with different values of $Q$ and  $P=0$ . 
The subfigure (II) is for  the case (2) with $Q=0.9$; $P=0$ and subfigure (III) is for the case (3) with $Q=0.9$; $P=0$.
The last subfigure (IV) gives a comparative plot for all three cases.}
\end{figure}

\begin{figure}[h]
\begin{tabular}{c c c c}
\includegraphics[width=0.5\linewidth]{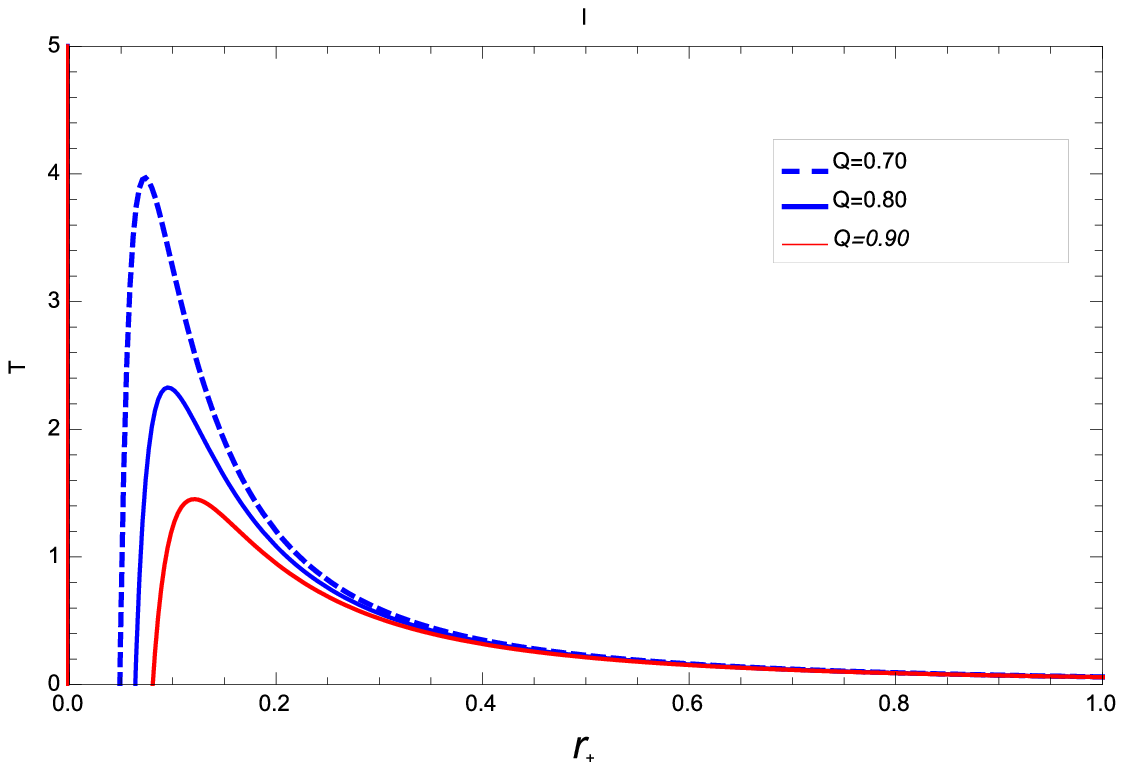}
\includegraphics[width=0.5\linewidth]{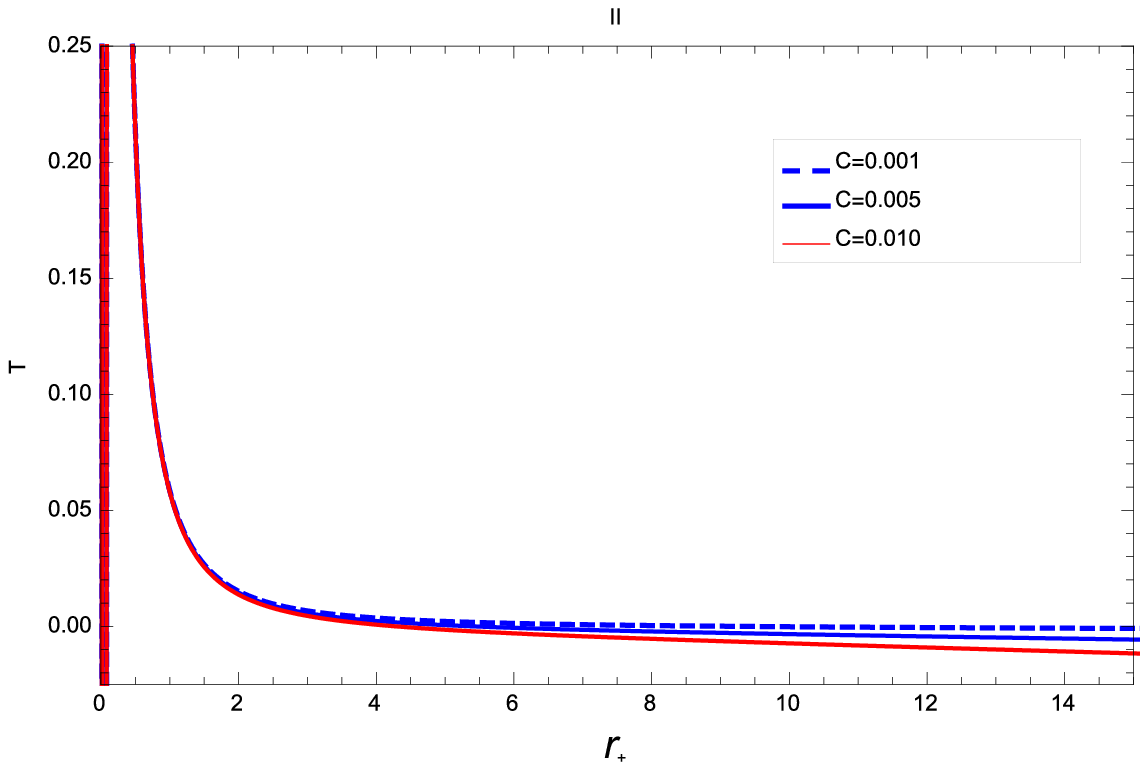}\\
\includegraphics[width=0.5\linewidth]{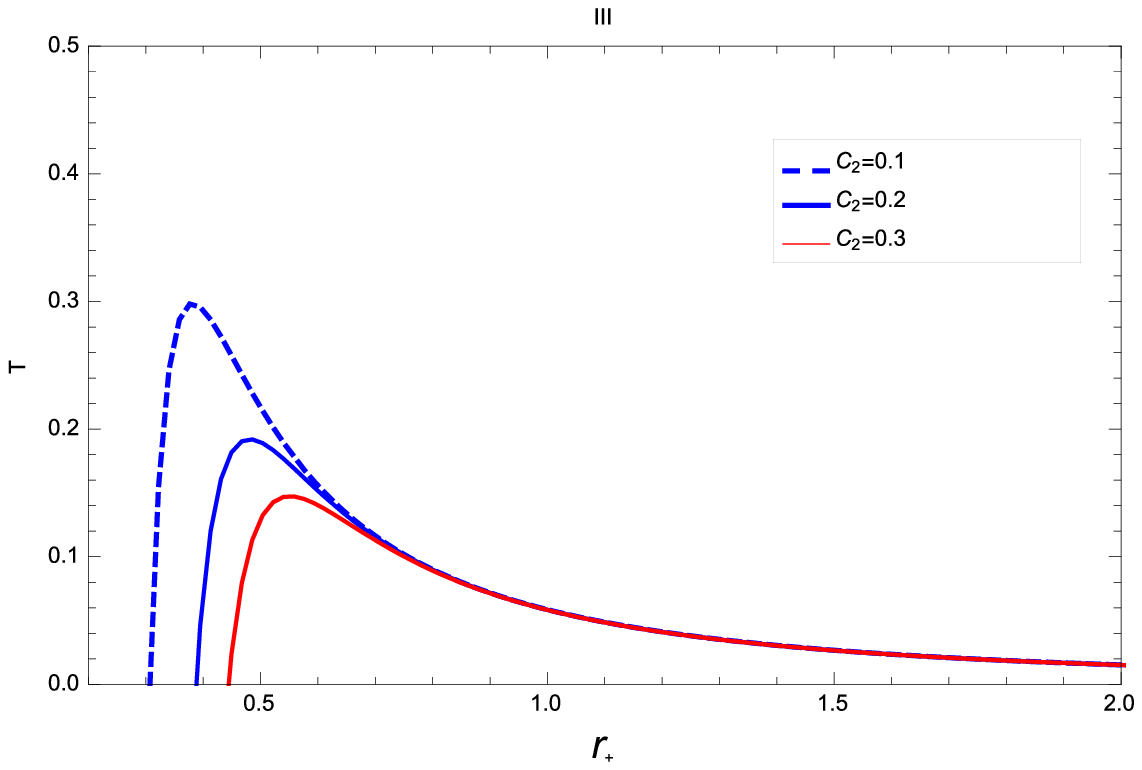}
\includegraphics[width=0.5\linewidth]{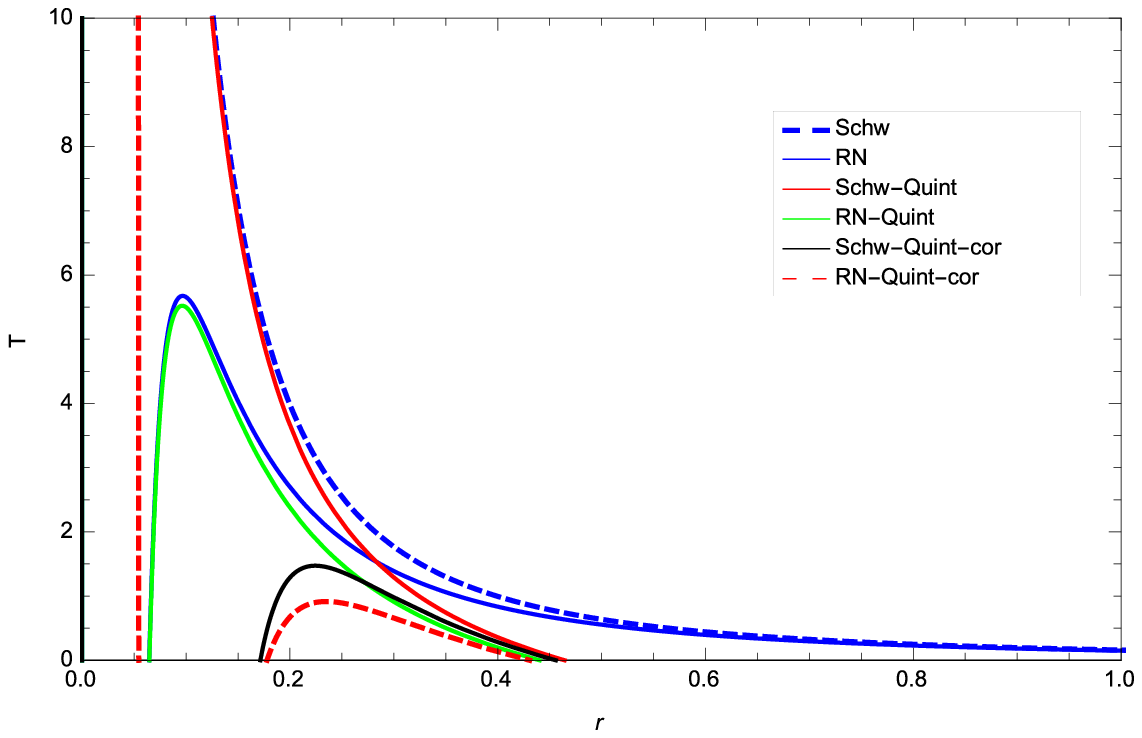}
\end{tabular}
\caption{\label{fig:RK1}The temperature with horizon radius $r_+$ is plotted in subfigures (I), (II) and (III) respectively
for the cases (1), (2) and (3). For the case (2) and case (3), $Q=0.9$; $P=0$. The last plot (IV) is a comparative plot.  }
\end{figure}

In this section, we will present the thermodynamical quantities in terms of horizon radius $r_+$ associated 
with the modified  R-N black hole solution  based on the general results in the previous section (\ref{E_Uni}). In this model, the 
modified R-N black hole solution (\ref{METRIC}) with metric components (\ref{eq:A_1}) and (\ref{metric_func}) is characterized by mass $M$, charge $Q$, cosmological constant term  $C$ and
correction term $C_2$. The mass of the modified RN black hole solution
is determined by using $A(r_+)=0$, and it turns out to be
\be
M=\frac{1}{6\pi r_{+} (8r_{+}^3-C_2)}\left[6r_{+}^3(Q^2 + 4\pi r_{+}^2-4 C \pi r_{+}^4)
-C_2(Q^2+2\pi r_{+}^2-3 C \pi r_{+}^4)-\frac{3}{8} C \pi r_{+} C_2^2\right]
\label{eqmp}
\ee
The variation of  mass of R-N black hole in unimodular gravity with horizon  radius $r_{+}$  is shown  in figure (\ref{fig:RK}).
Next, we calculate the thermodynamical quantities associated with the metric function (\ref{metric_func}). The Hawking temperature,
\be
T=\frac{\kappa}{2\pi},
\ee
where, the surface gravity $\kappa=-1/2\partial/\partial r\left(\sqrt{g_{tt}\,g^{rr}}\right)$ and  it is constant over the horizon.  
The Hawking temperature for modified RN black hole solution  is expressed as
\ba
\label{T}
T&&=\frac{\Big(1+\frac{C_2}{8r_+^3}\Big)}{384 \pi^2r_+^9}\Big[-96r_+^6\Big(Q^2-4\pi M r_+ +4\pi C r_+^4\Big)
+4r_+^3\Big(Q^2+6\pi r_+(2M-2r_+ +C r_+^3)\Big)C_2\nn\\&&\qquad\qquad+\Big(9Q^2+\pi r_+(-48M+14r_+ -3 C r_+^3)\Big)C_2^2+3\pi C r_+ C_2^2\Big].
\ea
The plots of temperature of the RN Black Hole for the cases (1), (2) and (3) are shown in figure ({\ref{fig:RK1}}).
Finally, we analyze the thermodynamic stability of the system which is related to the sign of the heat capacity ($H$). If
the heat capacity is positive ($H>0$), then the black hole is stable and if it is negative ($H<0$), the black hole
is said to be unstable. Here, the specific heat is given by
\be
H=\frac{\partial M}{\partial r_+}/\frac{\partial T}{\partial r_+}. \label{SpeHeat}
\label{cap}
\ee
Substituting the value of $M$ and $T$ in equation (\ref{SpeHeat}), we obtain  the specific heat,
\ba
H=\frac{512\pi r_+^{11}\Big(8+\frac{C_2}{r_+^3}\Big)^{2/3}\Big[48r_+^2\Big(Q^2+4\pi r_+^6(3 C r_+^2-1)\Big)-4 B_1 C_2
+B_2 C_2^2]}{(8r_+^3-C_2)^2\Big[768r_+^9 B_3 -192r_+^6C_2B_4+4r_+^3C_2^2B_5+2C_3^2B_6+27\pi r_+ C C_2^4 \Big]}
\label{cap}
\ea
where
\ba
&&B_1=5Q^2r_+^3+4\pi r_+^7(9ar_+^2-4),\qquad B_2=Q^2-2\pi r_+^2,\qquad B_3=3Q^2+8\p M r_+ +4\pi C r_+^4,\nn\\
&&B_4=Q^2-2\pi r_+(8M-4r_+ +C r_+^3) ,\quad B_5=169Q^2-4\pi r_+ (174M-34 r_+ +3 C r_+^3),\nn\\ && B_6=45Q^2
+\pi r_+(-216M+56 r_+ +81C r_+^3).
\ea
\begin{figure}[h]
\begin{tabular}{c c c c}
\includegraphics[width=0.5\linewidth]{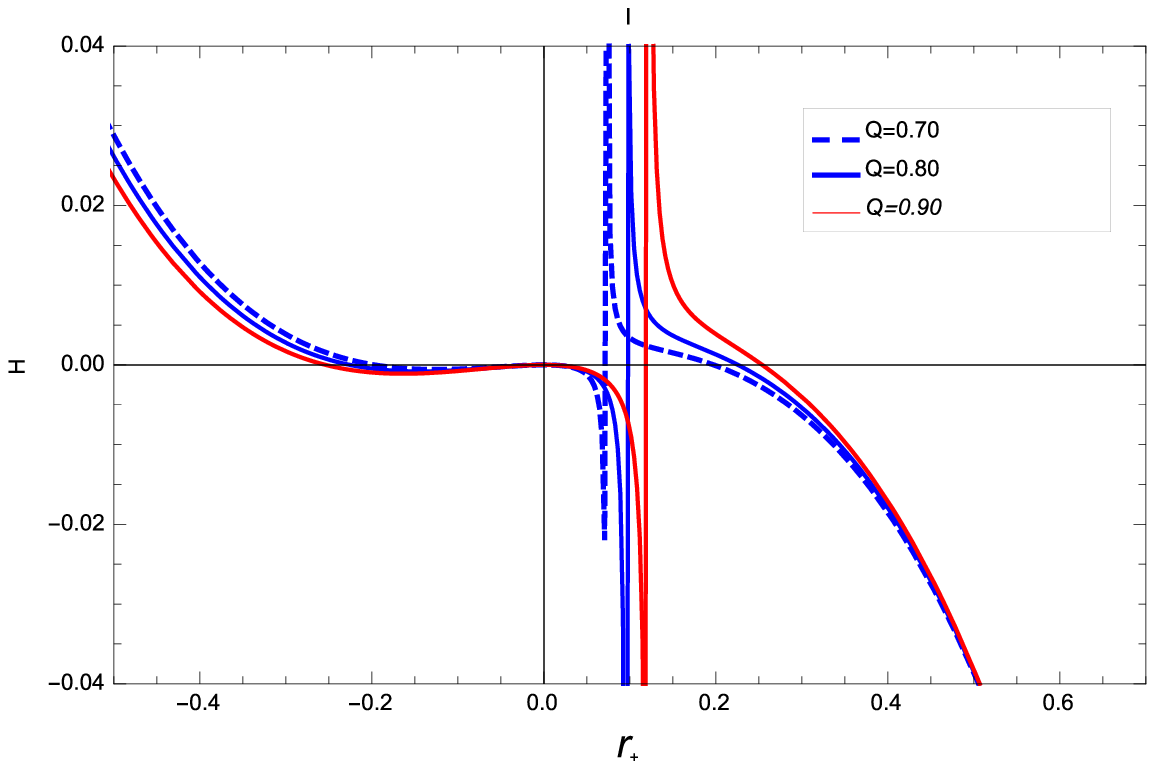}
\includegraphics[width=0.5\linewidth]{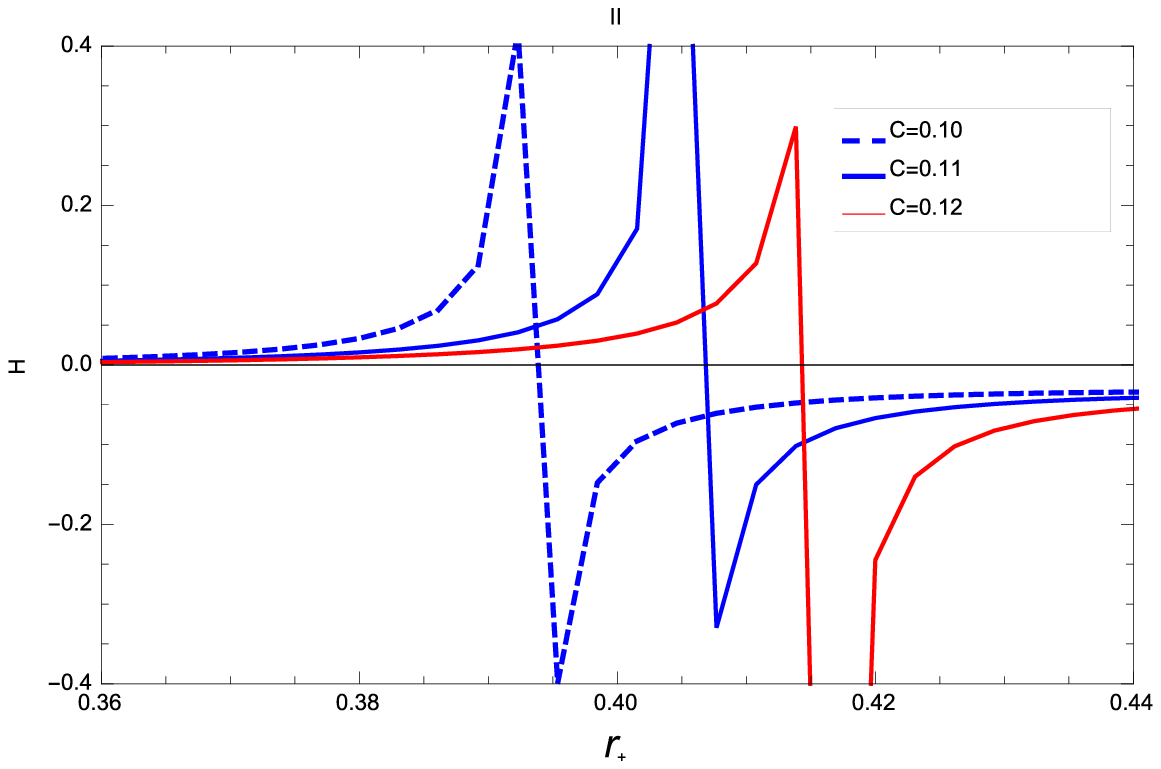}\\
\includegraphics[width=0.5\linewidth]{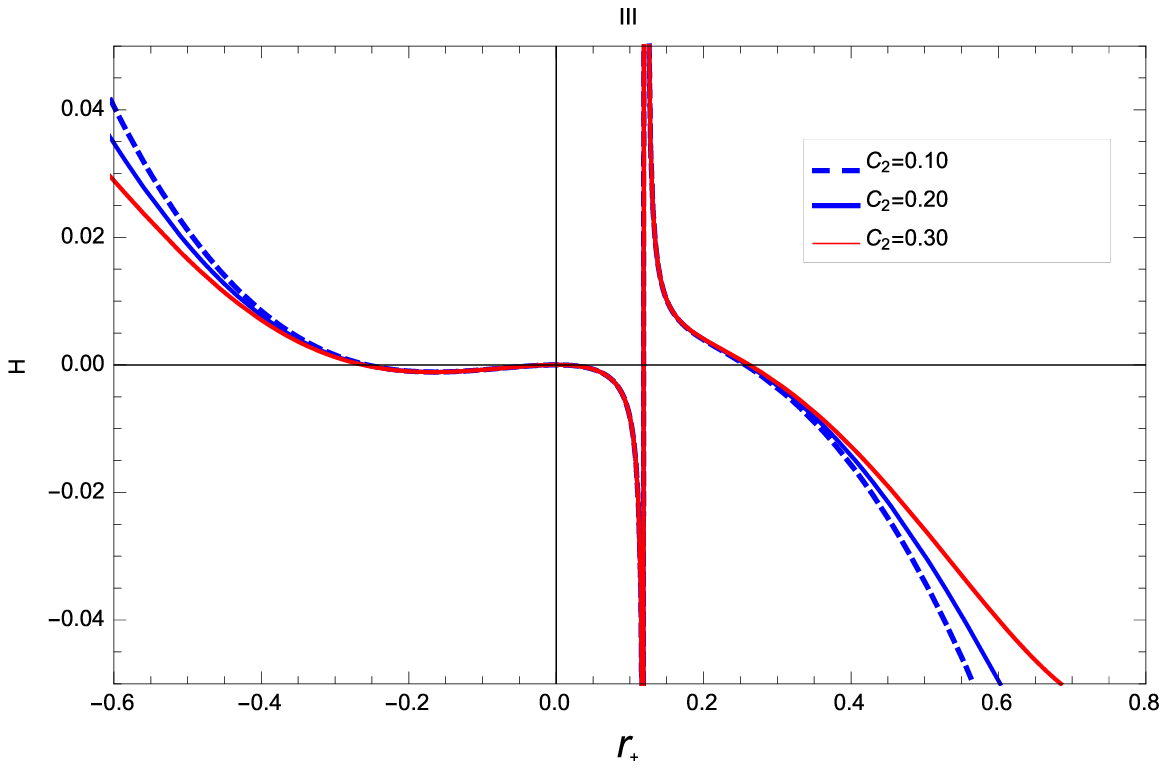}
\includegraphics[width=0.5\linewidth]{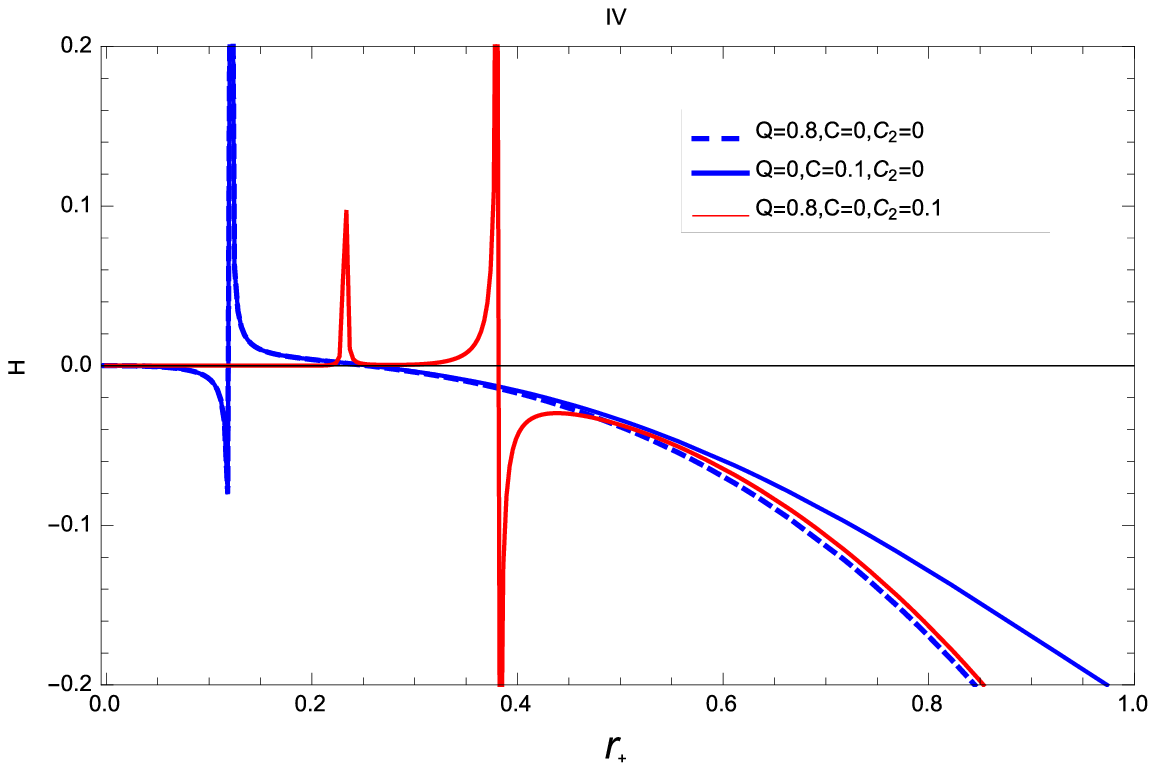}
\end{tabular}
\caption{\label{fig:RK2}The  specific heat with horizon 
radius $r_+$ is plotted in subfigures (I), (II) and (III) respectively
for the cases (1), (2) and (3). For the case (2) and case (3), $Q=0.9$; $P=0$. The last plot (IV) is a comparative plot.}
\label{figcap}
\end{figure}

 It is clear from Eq.~(\ref{cap}) that the heat capacity depends on the charge $Q$. When $Q \rightarrow 0$, $C \to 0$ and
 $C_2 \to 0$, one gets $C=-4\pi r^2_+$ which 
 means the black holes are thermodynamically unstable.  Next, we analyze the effect of cosmological constant on
 thermodynamical stability of black hole. We plot specific heat $H$ in Fig.~\ref{figcap} for different values of the
 constants $C$ for a fixed $Q$. It is seen that the heat capacity discontinuous at $r=r_C$ for each $C$ and for a given $Q$. We 
 observe that the heat capacity $H>0$ ($H<0$) for $r_+ < r_C$ ($r_+ >r_C$). Thus, the modified R-N black hole solution 
 is thermodynamically stable for $r_+<r_C$, as the 
 black hole has positive heat capacity and unstable for $r_+>r_C$ depicted in Fig. (\ref{figcap}).  


\section{Conclusion:}
We have solved for modified R-N metric  for the non rotating charged black hole in unimodular gravity. In  the calculation we have assumed that charge of black hole
is static and so the effect of magnetic field due to electric charge has not been considered.  We have found modification in the solution of R-N metric in the unimodular 
gravitational background. The correction is up to order of $ \sim 1/r^5$ to the leading order term corresponding to the charge of black-hole. Magnetic monopole
has been also taken into the account. We have shown that the leading order solution is same as in  R-N metric in usual gravity. We have also demonstrated that the components of
electromagnetic field tensor get modified by factor $\sqrt{A(r)B(r)}$ which is not exactly unity in unimodular gravity as in the case of
Schwarzschild solution. However
the leading order of $A(r)B(r)$ is unity. In the considered theory, the solution of metric defined in Eq. (\ref{METRIC})  depends
mainly on the four parameter Mass $M$, charge $Q$, cosmological constant $C_4$ (or $C$) and the correction terms due to the
constant $C_2$ which we assumed as very small. Both the corrections due to the constants $C_2$ and $C_4$ are the unimodular 
corrections and these are not present in standard general relativity. Considering all the features of unimodular gravity, we 
finally studied the thermodynamic properties  of the solution. One of interesting phenomena that could occur in R-N black hole 
is anti-evaporation  which explains about the primordial black hole. However, anti-evaporation at classical level needs
a $f(R)$-gravity \cite{Nojiri:2014jqa} or similar one; mimetic $f(R)$ gravity \cite{Oikonomou:2015lgy} and  it is also present
in Nariai space time at quantum level \cite{Niemeyer:2000nq}.  The studied unimodular gravity gives the Schwarzchild de-Sitter
space time if one sets $C_2=0$ with all charges zero. The metric solution with $C_2=0$ is same as what is studied in the
Ref. \cite{Nojiri:2014jqa}. However, in that Ref. \cite{Nojiri:2014jqa}, authors considered $f(R)$-theory and obtained certain
conditions to describe the anti-evaporation effect.  By looking the solution of metric, one may expect to have
anti-evaporation. However, this requires, in fact, the perturbation 
analysis to know the behavior of horizon radius, since the construction of  unimodular gravity is different from standard Einstein
gravity with cosmological constant. We hope to address these critical issues involving the charged black holes in unimodular theory
of gravity in our future works.

\section*{Acknowledgement}
Naveen K. Singh and Dharm Veer Singh are thankful to the D.S. Kothari postdoctoral fellowship of University Grant Commission, India for the
financial support under the fellowship number F.4-2/2006 (BSR)/PH/14-15/0034 and fellowship number BSR/2015-16/PH/0014 respectively. In addition Pankaj Chaturvedi is also thankful
to the Council of Scientific and Industrial Research (CSIR), India for the financial support under the Grant No. 09/092(0846)/2012-
EMR-I.

\bigskip
\noindent
\end{spacing}
\begin{spacing}{1}
\begin{small}

\end{small}
\end{spacing}

\begin{thebibliography}{unsrt}


\bibitem{Einstein} A. Einstein, in {\it The Principle of Relativity}, edited by
A. Sommerfeld (Dover, New York, 1952).
\bibitem{Anderson} 
J.L. Anderson and D. Finkelstein, Am. J. Phys. {\bf 39}, 901 (1971).

\bibitem{PSS} Petrosian, V., E.E. Salpeter and P. Szekeres, Astrophys. J. 
 {\bf 147}, 1222 (1967).
\bibitem{Shk} Shklovsky, I., Astrophys. J., 
 {\bf 150}, L1 (1967).
\bibitem{R_R} Rowan-Robinson, M., Mon. Not. R. Astron. Soc. 
 {\bf 141}, 445 (1968).
\bibitem{Perlmutter} S. Perlmutter {\it et al}., Astrophys. J. 
 {\bf 517}, 565 (1999).
\bibitem{Riess} A. G. Riess {\it et al}, Astron. J.
 {\bf 116}, 1009 (1998).
 \bibitem{Weinberg}
  S.~Weinberg,
  Rev.\ Mod.\ Phys.\  {\bf 61}, 1 (1989).
  
\bibitem{Jain:2011} P. Jain, P. Karmakar, S. Mitra, S. Panda and N. K. Singh, JCAP {\bf 05} 020 (2012).
\bibitem{Jain:2012gc} 
  P.~Jain, A.~Jaiswal, P.~Karmakar, G.~Kashyap and N.~K.~Singh,
  JCAP {\bf 1211}, 003 (2012)
  [arXiv:1109.0169 [astro-ph.CO]].
\bibitem{Cho:2014taa} 
  I.~Cho and N.~K.~Singh,
  Class.\ Quant.\ Grav.\  {\bf 32}, no. 13, 135020 (2015)
  doi:10.1088/0264-9381/32/13/135020
  [arXiv:1412.6205 [gr-qc]].
  
\bibitem{Bradenberger} 
  C.~Gao, R.~H.~Brandenberger, Y.~Cai and P.~Chen,
  JCAP {\bf 1409}, 021 (2014)
  doi:10.1088/1475-7516/2014/09/021
  [arXiv:1405.1644 [gr-qc]].
  
  
\bibitem{Basak} 
  A.~Basak, O.~Fabre and S.~Shankaranarayanan,
  arXiv:1511.01805 [gr-qc].
  
  \bibitem{Bamba:2016wjm} 
  K.~Bamba, S.~D.~Odintsov and E.~N.~Saridakis,
  arXiv:1605.02461 [gr-qc].
  
  \bibitem{Nojiri:2016plt} 
  S.~Nojiri, S.~D.~Odintsov and V.~K.~Oikonomou,
  arXiv:1605.00993 [gr-qc].
  
  \bibitem{Odintsov:2016imq} 
  S.~D.~Odintsov and V.~K.~Oikonomou,
  Astrophys.\ Space Sci.\  {\bf 361}, no. 7, 236 (2016)
  doi:10.1007/s10509-016-2826-9
  [arXiv:1602.05645 [gr-qc]].
  
  \bibitem{Nassur:2016yhc} 
  S.~B.~Nassur, C.~Ainamon, M.~J.~S.~Houndjo and J.~Tossa,
  arXiv:1602.03172 [gr-qc].
  
  \bibitem{Nojiri:2016ppu} 
  S.~Nojiri, S.~D.~Odintsov and V.~K.~Oikonomou,
  Class.\ Quant.\ Grav.\  {\bf 33}, no. 12, 125017 (2016)
  doi:10.1088/0264-9381/33/12/125017
  [arXiv:1601.07057 [gr-qc]].
  
  \bibitem{Nojiri:2016ygo} 
  S.~Nojiri, S.~D.~Odintsov and V.~K.~Oikonomou,
  Phys.\ Rev.\ D {\bf 93}, no. 8, 084050 (2016)
  doi:10.1103/PhysRevD.93.084050
  [arXiv:1601.04112 [gr-qc]].
  
 \bibitem{Nojiri:2015sfd} 
  S.~Nojiri, S.~D.~Odintsov and V.~K.~Oikonomou,
  JCAP {\bf 1605}, no. 05, 046 (2016)
  doi:10.1088/1475-7516/2016/05/046
  [arXiv:1512.07223 [gr-qc]]. 
  
\bibitem{abbassi1}
  Amir H. Abbassi and Amir M. Abbassi,
  \ Class.\ Quant.\ Grav.   {\bf 25}, 175018  (2008).

\bibitem{Fiol_Garriga}
  B.Fiol and J.Garriga,(Barcelona U. and  ICC,Barcelona U.), 
  JCAP\ {\bf 1008:015},\ 1475-7516 (2010). 
\bibitem{abbassi2}
  Amir H. Abbassi and Amir M. Abbassi,
  Annals\ Phys.\ {\bf 326},\ 1161-1173 (2011)
\bibitem{Bock}
 Robert Davis Bock,
 Int.\ J.\ Theor.\ Phys.\ {\bf 42}, 1835-1847 (2003)
\bibitem{Alvarez}
 Enrique Alvarez, (Madrid, Autonoma U.),
 JHEP \ {\bf 0503:002},(2005)
\bibitem{Alvarez_Faedo}
 Enrique Alvarez, Anton F. Faedo, (Madrid, Autonoma U.),
 Phys.\ Rev.\ D\ {\bf 76}, 064013 (2007). 
\bibitem{Farajollahi}
 Hossein Farajollahi, (Sydney U.) 
 Gen.\ Rel.\ Grav.\ {\bf 37}, 383-390 (2005).
\bibitem{Ng_Dam}
 Y.Jack Ng, H. van Dam, (North Carolina U.), 
 J.\ Math.\ Phys.\ {\bf 32}, 1337-1340 (1991). 
\bibitem{FGB}
David R. Finkelstein, Andrei A. Galiautdinov, James E. Baugh, (Georgia Tech),
 J.\ Math.\ Phys.\ {\bf 42}, 340-346 (2001). 
\bibitem{Earman}
J. Earman, (Pittsburgh U.), 
 Stud.\ Hist.\ Philos.\ Mod.\ Phys.\ {\bf 34}, 559-577 (2003). 
\bibitem{Mammadov}                   
Gulmammad Mammadov, 
Lecture Notes: Department of Physics,  Syracuse University, Syracuse, NY, USA,

\bibitem{Nojiri:2014jqa} 
  S.~Nojiri and S.~D.~Odintsov,
  Phys.\ Lett.\ B {\bf 735}, 376 (2014)
  doi:10.1016/j.physletb.2014.06.070
  [arXiv:1405.2439 [gr-qc]].
\bibitem{Oikonomou:2015lgy} 
  V.~K.~Oikonomou,
  Universe {\bf 2}, no. 2, 10 (2016)
  doi:10.3390/universe2020010
  [arXiv:1511.09117 [gr-qc]].
  \bibitem{Niemeyer:2000nq} 
  J.~C.~Niemeyer and R.~Bousso,
  Phys.\ Rev.\ D {\bf 62}, 023503 (2000)
  doi:10.1103/PhysRevD.62.023503
  [gr-qc/0004004].
  
\end{thebibliography}
\end{document}